\documentclass[pra,reprint,groupedaddress,secnumarabic,eqsecnum,showkeys,showpacs,raggedbottom,nofootinbib]{revtex4-1}

\usepackage{amsmath,amssymb,amsfonts,amsbsy,bm}   
\PassOptionsToPackage{caption=false}{subfig}      
\usepackage{graphicx,graphics,color,epsfig,subfig}
\usepackage{hyperref}                             
\usepackage[sort&compress]{natbib}                
\usepackage{color}                                

\usepackage{float}                                
\usepackage{MnSymbol}                              
\usepackage{amsthm}                                

\DeclareGraphicsExtensions{.pdf}    

\renewcommand{\a}{\alpha}
\renewcommand{\b}{\beta}
\renewcommand{\r}{\rho}
\newcommand{\s}{\sigma}
\newcommand{\m}{\mu}
\newcommand{\n}{\nu}

\renewcommand{\o}{\omega}
\renewcommand{\l}{\lambda}

\newcommand{\half}{\frac{1}{2}}

\newcommand{\bra}{\langle}
\newcommand{\ket}{\rangle}
\newcommand{\be}{\begin{equation}}
\newcommand{\ee}{\end{equation}}
\newcommand{\ba}{\begin{eqnarray}}
\newcommand{\ea}{\end{eqnarray}}
\newcommand{\dg}{\dagger}
\renewcommand{\aa}{\hat{a}}
\newcommand{\rin}{\rho_\mathsf{in}}
\newcommand{\rout}{\rho_\mathsf{ss}}
\newcommand{\mreg}{\mathsf{L}}
\newcommand{\mout}{\mathsf{L_{ss}}}
\newcommand{\uout}{U_{\mathsf{ss}}}
\newcommand{\pout}{\Pi_{\mathsf{ss}}}

\newcommand{\hout}{H_\mathsf{ss}}
\newcommand{\ph}{\hat{n}}
\newcommand{\hi}{{F}_{l}}
\newcommand{\LL}{\mathcal{L}}
\newcommand{\tl}{\mathcal{\hat{L}}}
\newcommand{\kket}{\rangle\hspace{-1mm}\rangle}
\newcommand{\bbra}{\langle\hspace{-1mm}\langle}
\newtheorem*{proposition}{Conserved quantity -- steady state correspondence}

\begin{document}

\title{Symmetries and conserved quantities in Lindblad master equations}
\author{Victor~V.~Albert}
\email[]{valbert4@gmail.com}
\affiliation{Departments of Applied Physics and Physics, Yale University, New Haven, Connecticut 06520, USA}
\author{Liang~Jiang}
\email[]{liang.jiang@yale.edu}
\affiliation{Departments of Applied Physics and Physics, Yale University, New Haven, Connecticut 06520, USA}
\date{\today}
\begin{abstract}

This work is concerned with determination of the steady-state structure of time-independent  Lindblad master equations, especially those possessing more than one steady state. The approach here is to treat Lindblad systems as generalizations of unitary quantum mechanics, extending the intuition of symmetries and conserved quantities to the dissipative case. We combine and apply various results to obtain an exhaustive characterization of the infinite-time behavior of Lindblad evolution, including both the structure of the infinite-time density matrix and its dependence on initial conditions. The effect of the environment in the infinite time limit can therefore be tracked exactly for arbitrary state initialization and without knowledge of dynamics at intermediate time. As a consequence, sufficient criteria for determining the steady state of a Lindblad master equation are obtained. These criteria are knowledge of the initial state, a basis for the steady-state subspace, and all conserved quantities. We give examples of two-qubit dissipation and {single-mode} $d$-photon absorption where all quantities are determined analytically. Applications of these techniques to quantum information, computation, and feedback control are discussed.

\end{abstract}
\maketitle

Environmental/reservoir interaction features rather prominently in the design,  engineering, and realization of quantum systems. Many models exist for simulating the environment, with the most prominent being the framework of GKS-L (Gorini-Kossakowsi-Sudarshan-Lindblad, or just \textit{Lindblad}) master equations \cite{Gorini1976a, *Lindblad1976}. Such master equations are valid only for specific Markovian environments (e.g. \cite{breuer, wisemanmilburn}). Nevertheless, Lindblad master equations continue to be implemented in a multitude of systems, lying in a nexus between quantum optics, quantum information, mesoscopic physics, and dynamical systems theory. We briefly list some notable works.

While system-environment interaction in the case of cavity (circuit) quantum mechanics usually consists of optical (microwave) photon loss, recent efforts have been to design the cavity such that other forms of dissipation can be realized. This can be done in order to control the state of either the qubit \cite{Murch2012,*Leghtas2013} or the cavity \cite{zaki}. There is much interest in designing dissipative mechanisms in other areas as well, e.g. trapped ions \cite{zoller_stabilizers} and optomechanics \cite{Tomadin2012}. Novel theoretical work has applied dissipation to topological systems, e.g. {a class of fermionic Lindblad systems \cite{bardyn}, }optical lattices \cite{Diehl2011, *Kraus2012}, atomic superfluids \cite{Bardyn2012}, the Creutz model \cite{Viyuela2012}, and the Haldane model \cite{Rivas2013}. Other prominent Hamiltonian systems have been studied with the addition of dissipation, e.g. Heisenberg/Hubbard spin chains \cite{prozenchains_prl2, *prozenchains_prl1, *prozenhubbard, popkov2013a, *Mendoza-Arenas2013, popkov2013, cai2013}, the Bose-Hubbard model \cite{Diehl2010}, the Ising model \cite{Foss-Feig2013}, and cold-atom systems \cite{Diehl2008, *Yi2012}. {A current topic in quantum control is the design of a Lindblad operator to obtain a desired, often exotic,  steady state \cite{pechen2006, *Kraus2008, *Ticozzi2009, *Wang2010, *pechen2011, *Ticozzi2012, Verstraete2009, schirmer} or steady-state property \cite{Sauer2013}. Recent developments are summarized in \cite{Muller2012} and refs. therein.}

Regarding open systems with multiple (degenerate) steady states, work has been spearheaded by the concepts of pointer states/subspaces \cite{zurek1981, *zurek1982, *zurek2003}, noiseless quantum codes \cite{zanardi1997, *zanardi1998a}, decoherence-free subspaces \cite{lidar1998, kempe2001, lidar2003}, and noiseless subsystems \cite{knill2000}. The latter two structures have been realized experimentally in several systems \cite{kwiat2000, *dalvit2000, *viola2001, *kielpinski2001} and continue to be a promising avenue for storing and processing quantum information. Proof-based works \cite{Ticozzi2008, robin} characterize these broad concepts as they relate to quantum information and quantum control. Recent efforts have begun studying transport properties in such degenerate Lindblad systems \cite{prozen, manzano2013}. Finally, there have been several pioneering mathematical works on generic Lindblad evolution operators \cite{Frigerio1977, *Spohn1977, *evans1977}, also known as \textit{quantum dynamical semigroups} (\cite{alicki_book, *rivas_book} and refs. therein), with research continuing in this domain \cite{Fagnola2008, *Dhahri2010, *Rouchon2013}.

The intractable literature on Lindblad systems begs the question of why this work is useful. While the properties of abstract Lindblad systems are garnering interest from physicists due to increasing ability to engineer previously un-physical Lindblad models, a gap in accessibility and nomenclature nevertheless remains (resonating with note 1.4 in \cite{carmichael1}). In the spirit of bridging this gap using physical intuition, this work points out the utility of symmetries and conserved quantities from ordinary quantum mechanics in the Lindblad formalism. We answer the following questions: {\it (1) How are symmetries and conserved quantities different in Lindblad systems when compared to unitary systems?} and {\it (2) Despite Lindblad evolution  being irreversible, what information from an initial state is preserved in the infinite-time limit?} In answering these questions, we apply previous results, develop our own when necessary, and provide physical examples. 

Sec.~\ref{sec1} addresses question (1), where we follow and elaborate on discussions of symmetries and/or conserved quantities from \cite{baum2,prozen}. In short, the correspondence between a continuous symmetries and conserved quantities (akin to Noether's theorem from field theory) is broken. The formulation here clarifies the relationship between symmetries and conserved quantities, provides conditions on when/how the two are related, and applies them to physical dissipative systems. Continuous and some discrete symmetries are discussed. The comparison to unitary evolution is made in order to make technical results of Lindblad theory accessible to an audience with only a background in quantum mechanics.

Sec.~\ref{sec2} begins to address question (2), whose full answer (given in Sec.~\ref{sec4}) is related to the structure of the \textit{limit set} \cite{teschl2012} (also \textit{set of asymptotic states} or \textit{invariant set}) of the Lindblad evolution operator, i.e., the set of density matrices evolving under a Lindblad operator in the infinite-time limit. The limit set consists of the set of steady states of the Lindblad operator and any surviving density matrices undergoing unitary rotations (oscillating coherences) induced by the Lindblad operator. Results used here in the determination of the limit set can be organized by the following chain of subsets:
\ba\nonumber
\text{Hamiltonians}\nonumber&\subseteq&\nonumber\\
\text{diagonalizable Lindblad operators 
        \cite{weidlich1971, briegel1993, *briegel1994, *barnett2000, buchleitner2002}}&\subseteq&\nonumber\\
\text{Lindblad operators 
        \cite{baum1, baum2, baumr, schirmer}}&\subseteq&\nonumber\\
\text{linear operators \cite{puri}}&\subseteq&\nonumber\\
\text{ generators of continuous dynamical systems
        \cite{teschl2012}}\nonumber
\ea
Since a Lindblad evolution operator is \textit{c}ompletely \textit{p}ositive (roughly speaking, Hermiticity-preserving; see B.3.3 of \cite{gregoratti2009} for a precise definition) and \textit{t}race-\textit{p}reserving at any given time, results regarding such \textit{CPTP maps} (also \textit{quantum channels}) \cite{robin} will also apply. When a Lindblad equation has more than one steady state, the information that is preserved in the infinite-time limit is dependent on the initial density matrix. A major utility of conserved quantities is to determine this information. This utility, described in Eqs. (\ref{basis}-\ref{rhoeq}), is in the form of a correspondence between steady states and conserved quantities that completely characterizes the dependence of the steady state on the initial state (i.e. when no oscillating coherences are present). This result can be derived from the generalized eigenvector decomposition of exponentials of linear operators (Eq. (10.23) in \cite{puri}; see also \cite{sokolov2006,brink2001}). It is proven here using properties of Lindblad operators and linear algebra. The correspondence is then extended to include oscillating coherences and combined with statements about CPTP operators \cite{robin} to provide an exhaustive characterization of the limit set of Lindblad master equations (Sec.~\ref{sec4}). This characterization complements results for general \cite{baum1, baum2, baumr, schirmer} and fermionic/bosonic \cite{prosen2008,prosen2010,Prosen2010b} Lindblad operators by explicitly determining the corresponding asymptotic state for arbitrary initial conditions. It is related to the discovery of a complete basis for a class of Lindblad operators \cite{weidlich1971} (later called a \textit{damping basis}) that has been utilized to solve Lindblad master equations in quantum optics \cite{briegel1993, *briegel1994, *barnett2000, buchleitner2002}. Previous efforts have attempted to extend this framework to the degenerate case \cite{jakob2004}, but an explicit mathematical demonstration was not provided. While one cannot always obtain a complete eigenvector basis for the entire space, we show that such a basis exists for elements of the limit set. Therefore, this procedure bypasses the need to consider dynamics in intermediate time. 

Sec.~\ref{sec3} presents Examples of qubit and oscillator (photonic) systems with the goal of revealing the physical significance of the information preserved in the steady state. The considered family of photonic master equations allows one to store the phase of a coherent state while qubit systems store coherences between certain Bell states. While the application in mind here is the preservation of quantum information, the results apply to all finite-dimensional and some physically relevant infinite-dimensional Lindblad models.

Sec.~\ref{sec4} characterizes the full limit set of Lindblad master equations, including both steady states and any unitary evolution induced in the infinite-time limit. We discuss how the tools developed here relate to quantum information and quantum control in Sec.~\ref{sec5} and provide an outlook in Sec.~\ref{sec6}.

\section{Unitary vs. dissipative systems}\label{sec1}

To better understand the effects of dissipation, it is worthwhile to compare it to unitary evolution. Since we consider decoherence, we will discuss both systems from the point of view of density matrices living in an $N^2$-dimensional \textit{matrix} Hilbert space $\mreg$ (Liouville space \cite{ernstnmr, tarasov_book}) with inner product $\bbra\r|\s\kket = \text{Tr}\{\r^\dagger\s\}$ for $\r,\s\in\mreg$. We emphasize that $\r$ are matrices and restrict a modified bra-ket notation to the appendices. Throughout the paper, $\rin$ and $\rout$ are states in $\mreg$, capital symbols are operators in $\mreg$, calligraphic symbols (e.g. $\LL $) are (super-)operators on $\mreg$, indexed lower-case symbols are coefficients, and bosonic operators are $[\aa,\aa^\dg]=I$ and $\ph=\aa^\dg\aa$. Greek indices enumerate the steady-state subspace.

Unitary systems evolve in time under a one-parameter continuous group generated by the system Hamiltonian $H$. Dissipative systems evolve in time under a one-parameter \textit{semi}group $\{e^{\LL t},t\geq0\}$ generated by the Liouvillian $\LL$. Since the time-evolution operator $e^{\LL t}$ is no longer unitary, a state may undergo additional trajectories associated with negative real eigenvalues (decay) and complex pairs of eigenvalues with negative real parts (spiraling; see Fig. \ref{f1}) \cite{baum2}. Since it is not Hermitian, $\LL $ may no longer be diagonalizable. Naturally, time evolution under $\LL$,
\begin{equation}\label{lind}
\dot{\r}=\LL (\r)=-i\left[H,\r\right]+\sum_{l=1}^{N^{2}-1}2\hi\r \hi^{\dg}-\hi^{\dg}\hi\r-\r \hi^{\dg}\hi
\end{equation}
is generally not reversible (formally, $\LL$ generates a contraction semigroup \cite{alicki_book, *rivas_book}). In the above equation, $\{\hi\}_{l=1}^{N^2-1}$ are the dissipation-inducing ``jump'' operators and each $\hi$ can depend on a parameter. Examining the expectation value of an operator $J$ ($\text{Tr}\{J\r\}$), one can use the cyclic property of the trace to obtain its equation of motion (the Heisenberg picture)
\begin{equation}\label{adj}
\dot{J}=\LL ^{\dagger}(J)=i\left[H,J\right]+\sum_{l=1}^{N^{2}-1}2\hi^{\dg}J \hi-\hi^{\dg}\hi J\--J \hi^{\dg}\hi
.\end{equation}

\subsection{Steady states \& oscillating coherences}\label{sec11}

Steady-state density matrices are constructed out of eigenvectors of $\LL $ whose eigenvalue is zero (i.e. the kernel of $\LL $).\footnote{Note that in the case when $\LL $ has no zero eigenvalues, the steady state is unique \cite{baum1}.} Those eigenvectors consist of two types of elements: steady states and steady-state coherences. Adapting discussion from \cite{baumr}, we first describe them for unitary evolution (noting that these definitions are basis-dependent).

When there are no degeneracies, \textit{steady states} (also \textit{stationary states} or \textit{fixed points}) of a unitary time evolution operator are constructed out of pure states (i.e. projections) that commute with the Hamiltonian $H$. The set $\{|E_{i}\ket\bra E_{j}|\}_{i,j=1}^{N}$ with $H|E_i\ket = E_i |E_i\ket$ is a basis for the space of operators $\mreg$. There will be at least $N$ steady-state basis elements since all $|E_i\ket\bra E_i|$ commute with $H$. If there is additionally a degeneracy between levels $k\neq l$, then $|E_k\ket\bra E_l|$ will be a {\it steady-state coherence} between steady states $|E_k\ket\bra E_k|$ and $|E_l\ket\bra E_l|$. In that case, it is easy to see that density matrices written with $|E_{\iota}\ket\bra E_{\iota'}|$ (where $\iota,\iota'=k,l$) can be unitarily manipulated without leaving the steady-state subspace.
Finally, any coherence $|E_{m}\ket\bra E_{n}|$ with $m\neq n$ and no degeneracy ($E_m\neq E_n$) will be an \textit{oscillating coherence}, i.e., will rotate with a phase $i(E_n - E_m)t$. 

Using the same intuition, we now discuss the three aforementioned italicized concepts for the dissipative case. Due to decay, dissipative systems may have \textit{less than} $N$ steady states and those states may not be pure. However, a finite dissipative system will have at least one steady state (e.g. Prop. 5 in \cite{baum2}) and many infinite systems with physical relevance and/or reasonable finite limits do also. In other words, the dimension of the steady-state subspace $\mout\subseteq\mreg$, the eigenspace of eigenvalue zero of $\LL $, is between 1 and $\dim\{\mreg\}=N^2$.\footnote{While all steady states are elements of $\mout$, not all elements of $\mout$ are states since the set of density matrices is not closed under addition.} There may be steady-state coherences (also stationary phase relations \cite{baum2}) under Lindblad evolution when $\dim\{\mout\}\geq4$. The space $\mout$ is in general determined by both $H$ and $\hi$ from Eq.~(\ref{lind}).

Just as with unitary evolution, oscillating coherences are induced by unitary rotations on steady-state coherences. Two important statements about oscillating coherences are in order: (1) they are induced only by a Hamiltonian part of $\LL$ \cite{baum2} and (2) not all Hamiltonians induce them. Statement (1) implies that oscillating coherences can be rotated out by going into the rotating frame of the Hamiltonian that induces them. Thus, we will focus on steady states and steady-state coherences throughout the paper, mentioning oscillating coherences only in the general characterization in Sec.~\ref{sec4} and Appendix \ref{ap3}. Regarding statement (2), the forthcoming Example~\ref{sec33} is of a system with a Hamiltonian and no oscillating coherences.

\begin{figure}
\centering
\includegraphics[width=0.8\linewidth]{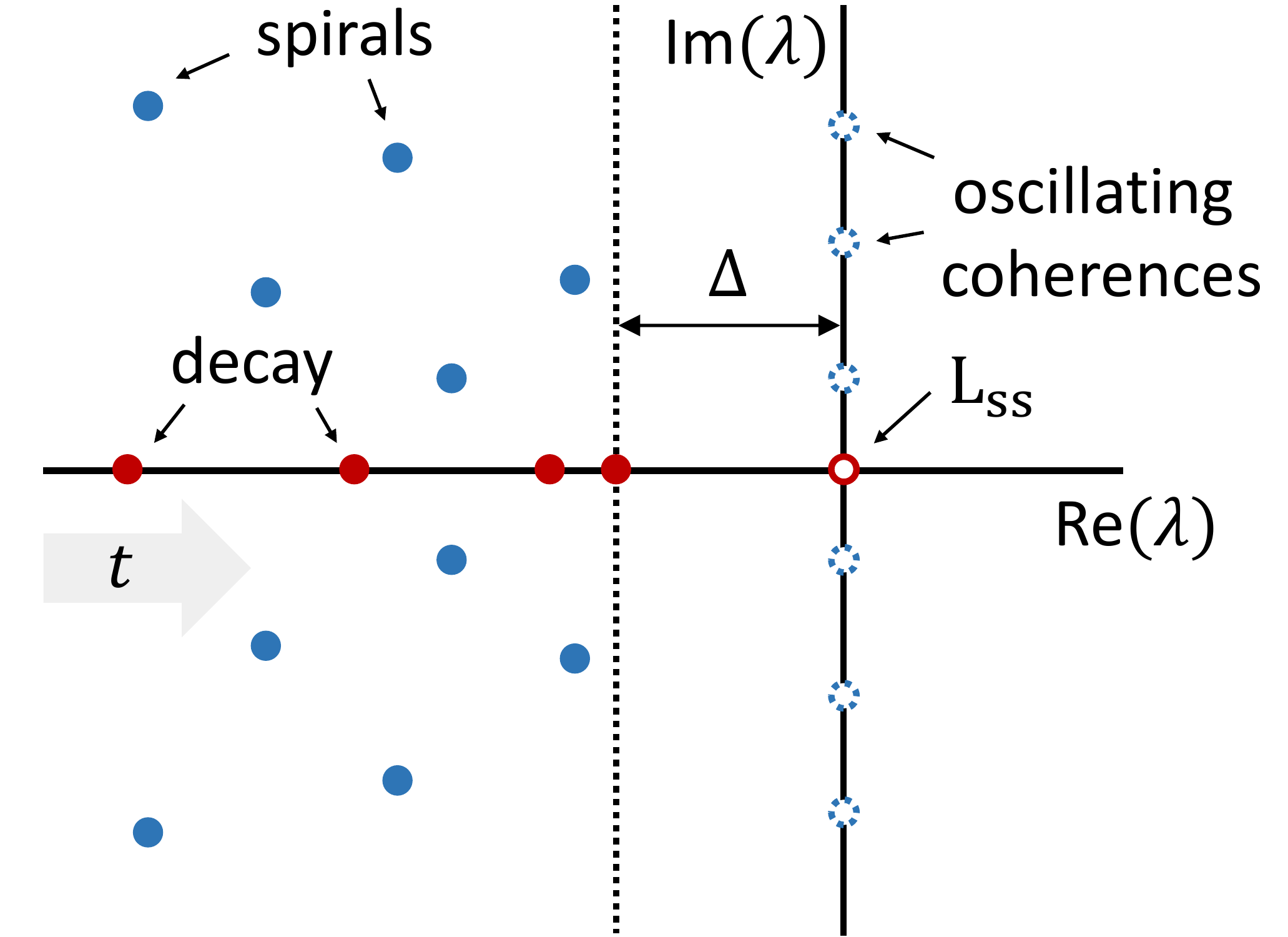}
\caption{(color online) While $\LL $ may not be diagonalizable, one can still obtain information about the dynamics by observing its  eigenvalues $\lambda$ in the complex plane. All eigenvalues lie on the non-positive plane and non-real eigenvalues exist in complex conjugate pairs (blue). Solid circles depict eigenvalues that cause a loss of portions of the initial density matrix (decay in red and spirals in blue). Unfilled circles represent $\lambda$ whose eigenstates survive in the infinite-time limit (steady states/steady-state coherences in red and oscillating coherences in dashed blue).  The grey arrow depicts time evolution towards the infinite-time $\lambda$. The value $\Delta$ is the \textit{dissipation/spectral gap} \cite{prosen2008, cai2013} -- the slowest non-zero rate of convergence toward the infinite-time states.}
\label{f1}
\end{figure}

The above introduction can be summarized in Fig. \ref{f1}, where a possible spectrum of $\LL$ is plotted. Unfilled circles completely characterize unitary evolution while the addition of negative real parts generalizes the analysis to Lindblad equations. 

\subsection{Steady-state structures}\label{sec111}

For dissipative systems with no oscillating coherences, an \textit{initial density matrix} $\rin\in\mreg$ will evolve under $e^{\LL t}$ in the infinite time limit into the corresponding \textit{asymptotic} or \textit{steady-state density matrix} $\rout\in\mout$ (with $\rout$ in general depending on $\rin$):
\begin{equation}
e^{\LL t}:\rin\xrightarrow{t\rightarrow\infty}\rout.
\end{equation}
If one assumes that the exponential convergence is fast compared to all other relevant timescales (i.e., the dissipation gap $\Delta$ from Fig. \ref{f1} is large; details in Sec. \ref{sec5}), then one can interpret $\LL $ as a black box that transforms $\rin$ into $\rout$. 
Below are five examples of structures of $\rout$ that can occur in dissipative systems:
\begin{eqnarray}\label{exam}
\begin{array}{c}
\begin{array}{ccc}
\begin{pmatrix}1\,\,\,\,\, & \cdot\,\,\,\,\, & \cdot\\
\cdot\,\,\,\,\, & \cdot\,\,\,\,\, & \cdot\\
\cdot\,\,\,\,\, & \cdot\,\,\,\,\, & \cdot
\end{pmatrix}, & \begin{pmatrix}\frac{1}{3}\,\,\,\,\, & \cdot\,\,\,\,\, & \cdot\\
\cdot\,\,\,\,\, & \frac{1}{3}\,\,\,\,\, & \cdot\\
\cdot\,\,\,\,\, & \cdot\,\,\,\,\, & \frac{1}{3}\,
\end{pmatrix}, & \begin{pmatrix}\r_{00} & \cdot & \cdot\,\,\,\\
\cdot & \r_{11}\,\,\, & \cdot\,\,\,\\
\cdot & \cdot & \cdot\,\,\,
\end{pmatrix}\end{array}\\
\begin{array}{cc}
\begin{pmatrix}\r_{00} & \r_{01} & \cdot\\
\r_{10} & \r_{11} & \cdot\\
\cdot & \cdot & \cdot
\end{pmatrix}, & \begin{pmatrix}\r_{00} & \r_{01} & \cdot\\
\r_{10} & \r_{11} & \cdot\\
\cdot & \cdot & \r_{22}
\end{pmatrix}\end{array}
\end{array}
\end{eqnarray}

In the above list, dots indicate the portions of $\mreg$ unavailable to $\mout$ and do not represent single zeroes in a matrix. The dimension of $\mout$ is 1, 1, 2, 4, and 5 in the respective structures. Of course, the complex coefficients $\rho_{\m\n}$ obey the well-known properties that make $\rout$ a density matrix: $\text{Tr}\{\rout\} = 1$, $\r_{\n\m}=\r_{\m\n}^{\star}$, and (by the Cauchy-Schwarz inequality) $\r_{\m\m}\r_{\n\n}\geq|\r_{\m\n}|^{2}$. The number of independent real parameters is thus $\dim\{\mout\}-1$. Since they contain independent parameters, the third through fifth structures are \textit{information preserving} \cite{robin}. What is meant by {\it information} in this work is simply any parameters that characterize $\rin$.

\textit{First structure}: unique pure steady state. A relevant physical system is single photon loss in a non-driven zero-temperature cavity,
\be\label{sp}
\LL (\rin) = 2\aa \rin \aa^\dg -\aa^\dg \aa \rin -\rin \aa^\dg \aa,
\ee
with $[\aa,\aa^\dg]=1$. While this example is infinite, one can apply the techniques here rigorously for any reasonable finite photon number limit. The damping sets the steady state to be the vacuum $|0\ket\bra0|$ (where $|n\ket$ is a Fock state), and the addition of a resonant drive/pump (e.g. Sec.~9.1 in \cite{klimov_book}) will set the steady state to be a coherent state independent of $\rin$. As a more general fact, irrespective of the form of the drive $H$, the steady state of a system with dissipation governed by Eq. (\ref{sp}) will always be unique and dependent only on the parameters in $\LL$ (Sec.~IV-B in \cite{schirmer} or Eq. (27) in \cite{Prosen2010b}).

\textit{Second structure}: unique mixed steady state. In this and the previous structure, the state is said to be \textit{attractive} (i.e. all initial states converge to it). A related example is the damped harmonic oscillator (Eq. (6.1.3) in \cite{zoller_book}). Its steady state is a Boltzmann distribution, i.e.,  a (thermal) equilibrium steady state. Relevant examples of unique \textit{non-equilibrium} steady states exist in open Hubbard and Heisenberg spin chains \cite{prozenchains_prl2, *prozenchains_prl1, *prozenhubbard} {as well as models of photosynthetic transport} \cite{manzano2013}.

\textit{Third structure}: steady state is a mixed state with $\r_{\m\m}$ dependent on $\rin$ (Example~\ref{sec31} or Sec.~5.2 in \cite{baum2}). This is an example of a classical bit (with $\r_{\m\m}$ being probabilities of a ``0'' or a ``1'') or, alternatively, a \textit{pointer basis} made up of two pointer states \cite{robin}. While there are infinitely many possible states due to the degree of freedom in $\r_{\m\m}$, the dimension of the steady-state space is just two.

\textit{Fourth structure:} qubit steady state, i.e., a steady-state coherence develops between two of the $\r_{\m\m}$ (e.g. Sec.~5.3 in \cite{baum2}). The steady-state subspace is four-dimensional and $\rout$ can be expanded in terms of the four basis matrices $M_{\m\n}$ (with $\m,\n=0,1$). It is important to note that $M_{\m\n}$ do not have to be of the form $|\mu\ket\bra\nu|$. All $M_{\m\n}$ may share a non-trivial common matrix factor $T$ such that  each $M_{\m\n}$ is unitarily equivalent to $|\m\ket\bra\n|\otimes T$ \cite{baum2, robin}. This tensor product structure occurs if, for example, the jump operators $\hi\in \mreg$ can be unitarily decomposed into $I\otimes f_l$, where $I$ is the identity on the space of $|\m\ket\bra\nu|$ and $f_l$ is a jump operator on the space of $T$ \cite{kempe2001}. Whenever $T$ is two-by-two or greater, the above is an example of a \textit{noiseless subsystem} (NS; also decoherence-free subsystem) \cite{knill2000, kempe2001}. For a trivial $T$, this is a \textit{decoherence-free subspace} (DFS) \cite{lidar1998, lidar2003}. In Sec.~\ref{sec3}, we provide specific manifestations: Examples~\ref{sec32}, \ref{sec34}, and \ref{sec35} are DFS's while Example~\ref{sec33} is an NS.

\textit{Fifth structure}: $\mout$ is direct sum of a two-by-two and a one-by-one space. This is an example of a \textit{hybrid quantum memory} consisting of a classical and a quantum bit \cite{kuperberg2003}. Note that steady states in the last three structures can be either pure or mixed, depending on $\rin$. This example is most representative of the general structure of a steady state -- a matrix of blocks of varying sizes with each block sharing a potentially non-trivial matrix factor (Thm. 7 in \cite{baumr} or Thm. 5 in \cite{robin}). This general structure will become relevant in Example~\ref{sec33} and Sec.~\ref{sec4}.

\subsection{Symmetries \& conserved quantities}\label{sec12}

 In a unitary system, an (explicitly time-independent) observable $J=J^\dg$ is a conserved quantity (i.e. constant of motion) if and only if  it commutes with the Hamiltonian (e.g. angular momentum of the hydrogen atom). In the spirit of Noether's theorem, one can then generate a continuous symmetry $U= \exp(i\phi J)$ (for real $\phi$) that leaves the Hamiltonian invariant. There is thus the following set of equivalent statements for continuous symmetries in unitary evolution (with one-sided arrows depicting an ``if-then'' statement, two-sided arrows depicting ``iff,'' and the dot being total time derivative):
\begin{equation}\label{symreg}
\begin{array}{rcl}
 & [J,H]=0\\
\Neswarrow &  & \Nwsearrow\\
\dot{J}=0~ & \Leftrightarrow & ~U^{\dg}HU=H
\end{array}
\end{equation}

A conserved quantity in dissipative systems is one where $\LL ^\dagger(J)=0$. One needs to introduce the adjoint representation \cite{ernstnmr, tarasov_book} to discuss symmetries on the super-operator level:
\be
U^{\dg}\hi U=\mathcal{U}^\dg(\hi).
\ee
The super-operator $\mathcal{U}=\exp(i\phi\mathcal{J})$, where $\mathcal{J}=\mathcal{J}^\dg$ is the super-operator analogue of $J$. The precise relation between $J$ and $\mathcal{J}$ is in Eq. (\ref{jjj}) and we will consider only super-operators $\mathcal{U}$ which can be written in terms of a $J$ on the operator level. Using this notation, one can map an analogous set of statements for dissipative systems:
\begin{equation}\label{dissym}
\begin{array}{c}
~~[J,H]=[J,\hi]=0~~\forall l\\
\begin{array}{rcl}
\Swarrow~~~~~~~ &  & ~~~~~\Searrow\\
\dot{J}=\LL ^\dagger(J)=0~~~~~~ &  & ~~~~\mathcal{U}^{\dg}\LL \mathcal{U}=\LL 
\end{array}
\end{array}
\end{equation}
Four arrows are lost.  First, there exist conserved quantities which do not  commute with everything in $\LL $ but are conserved ``as a whole'' (see Example~\ref{sec35}).\footnote{While our definition of a conserved quantity matches the invariant observable of \cite{baum2}, it is more specific that the strong symmetry of \cite{prozen} precisely due to conservation not implying commutation. What this work refers to as a symmetry is called a dynamical symmetry in \cite{baum2} and a weak symmetry in \cite{prozen}. } Second, the $U$ (and therefore $\mathcal{U}$) generated by such quantities are not always symmetries of the system. Third, a symmetry generator $J$ does not have to be a conserved quantity. Fourth, a symmetry generator does not have to commute with everything in $\mathcal{L.}$ The third and fourth points stem from conservation being on the super-operator level, i.e., stemming from a $\mathcal{J}$ and not necessarily a $J$. Examples of all four cases and conditions on when $J$ is a symmetry for simple $\mathcal{L}$ are in Appendix \ref{ap1}. Finally, note that the identity $I$, while not necessarily proportional to a steady state, is always conserved (since $\LL $ is trace-preserving).

As seen above, symmetries and conserved quantities are generally independent in dissipative systems (and overlapping when $J$ commutes with everything in $\LL $). Since it seems that adding dissipation only reduces the utility of a symmetry-based analysis, it begs to question what conserved quantities and symmetries are useful for. In short, conserved quantities correspond to the $\rho_{\m\n}$ from Eq. (\ref{exam}) while symmetries can be used to block-diagonalize $\LL $ and perform unitary transformations on $\rout$. We discuss these uses below.

\section{Utility of symmetries and conserved quantities}\label{sec2}

\subsection{Conserved quantities}\label{sec21}

In duality to elements of $\mout$ being eigenmatrices of $\LL $ with right eigenvalue zero, conserved quantities $J$ are right eigenmatrices of $\LL ^\dg$ (or, alternatively, adjoints of the left eigenmatrices of $\LL $) with eigenvalue zero \cite{baum2}. Note that in unitary systems, the sets of left and right eigenmatrices are identical and this analysis is not necessary. The number of linearly independent (not always Hermitian) $J$ is the same as the dimension of $\mout$ (Sec.~2.2.3 in \cite{vaughn_book}). Moreover, there is a correspondence between the steady state basis elements and the conserved quantities. We state the result below, with proof in Appendix \ref{ap2}. Oscillating coherences have a similar relation, proved in Appendix \ref{ap3}.

\begin{proposition} Assume a Lindblad system $\LL $ has no purely imaginary eigenvalues and let $\{M_{\mu}\}_{\mu=1}^{D}$ be an orthonormal basis for the $D$-dimensional steady-state subspace $\textnormal{$\mout$}\subseteq \mreg$. Then, corresponding to any \textnormal{$\rin\in\mreg$},
\be\label{basis}
\textnormal{$\rout$}=\lim_{t\rightarrow\infty}e^{\LL t}(\textnormal{$\rin$})=\sum_{\mu=1}^{D}\r_{\mu}M_{\mu}
\ee
and there exist $D$ linearly independent conserved quantities $J_{\mu}$ such that
\be\label{rhoeq}
\textnormal{$\r_{\mu}=\text{Tr}\{J_{\mu}^{\dg}\textnormal{$\rin$}\}$}.
\ee
\end{proposition}

Once again note that each $M_{\mu}$ need not be a pure state projection (see Example \ref{sec33}), a density matrix, or even Hermitian. In this convention, they should be thought of as vectors: $\text{Tr}\{M_{\m}^\dg M_{\n}\}=\delta_{\m\n}$ and $\text{Tr}\{M_{\m}\}$ is not always 1. However, the $J_\m$ are normalized such that $\rout$ is a density matrix. The $M_\m$ can be thought of as the basis elements of each of the entries in Eq. (\ref{exam}) and interpreted as independently collecting information from $\rin$ without  exchanging information with each other \cite{baum2}. In unitary systems, $J_\m=M_\m^\dg$. To outline the proof for dissipative systems, since $\mathcal{L^\dg}(J_\mu)=0$ and thus
\be
\text{Tr}\{J_\mu^\dagger\rin\}=\text{Tr}\{J_\mu^\dagger\rout\},
\ee
the $J_{\mu}$ can be arranged and normalized in such a way that each one will reveal the contribution of $M_\mu$ to $\rout$. They can be thought of as Lindblad analogues of pointer observables \cite{zurek1981, *zurek1982, *zurek2003}. In the trivial case when the steady state is unique [first structure in Eq. (\ref{exam})], no information about $\rin$ is preserved and the identity is the unique conserved quantity.
  
 The above correspondence has exclusively utilized the \textit{vector} nature of both $M_\m$ and $J_\m$ and the additional property that both $M{_\m}$ and $J_\m$ are $\dg$-closed ($\{M_{\m}^{\dg}\}=\{M_{\m}\}$ and similarly for $\{J_\m\}$). As \textit{matrices}, conserved quantities form useful Lie algebras in unitary systems (e.g. angular momentum $su(2)$ Lie algebra for the Hydrogen atom). Unfortunately, due to the presence of decay, the $J_\mu$ may not form a Lie algebra (e.g. Sec.~5.3 in \cite{baum2}). A much simpler picture is obtained when decay is removed and $J_\m$ are restricted to $\mout$. Letting $\pout$ be the projection onto $\mout$, the set of $j_\m\equiv \pout(J_\m)$ \textit{does} form a Lie algebra.\footnote{Since $e^{\LL t}$ is a completely positive trace-preserving map for any $t$, the results of \cite{robin} regarding invariant spaces apply. The result states that $j_\m$ form a matrix algebra, a vector space of matrices closed under $\dagger$ and multiplication. A Lie algebra can be built out of a matrix algebra by simply adding the commutation operation \cite{heckman}.}  The structure of the set of $j_\m$ replicates (but is not identical to) the block-diagonal structure of $\mout$ (Thm. 5-iii in \cite{robin}), thereby relating steady states to conserved quantities in another way. Going backwards, $e^{\LL ^\dg t}j_\m\rightarrow J_\m$ as $t\rightarrow\infty$ \cite{baum2}. A more convenient way to determine $J_\m$ for finite systems is simply to find the nullspace of $\LL ^\dg$.

\subsection{Symmetries}\label{sec22}

As mentioned in Eq. (\ref{dissym}), a (global) continuous symmetry $U$ is a unitary operator whose corresponding super-operator $\mathcal{U}=e^{i\phi\mathcal{J}}$ is such that $\mathcal{U}^\dg \LL  \mathcal{U}=\LL $, or equivalently $[\mathcal{J},\LL ]=0$. It is therefore easy to see that both $\mathcal{U}$ and $\mathcal{U}^\dg$ are symmetries of both $\LL $ and $\LL ^\dg$. To state in a different way, $\mathcal{U}$ commutes with time-evolution generated by $\LL $,
\be
e^{\LL t}(U^{\dg}\rin U)=U^{\dg}e^{\LL t}(\rin)U
,\ee
for any $\rin\in\mreg$. Examples of symmetries include any $U$ such that $UHU^\dg = H$ and $U\hi U^\dg = e^{i\phi_{l}}\hi$ \cite{bardyn} or   any permutations among the jump operators  $\hi$ from Eq. (\ref{lind}) that leave $\LL$ invariant \cite{prozen}. The Liouvillian can be block-diagonalized by  $\mathcal{U}$ in the same way that a Hamiltonian can be block-diagonalized by $U$ (with each block corresponding to an eigenvalue of $U$). Symmetries can thus significantly reduce computational cost, with the additional complication that the blocks of $\LL $ may not be further diagonalizable.  However, symmetries by themselves \textit{do not} determine the dimension of $\mout$ because some blocks may contain only decaying subspaces and no steady states. Diagonal parts of $\rin$ will always be in blocks with steady states since the trace is preserved. For a unitary $U$ such that $[U,H]=[U,\hi]=0$, $\dim\{\mout\}$ will be at least as much as the number of distinct eigenvalues of $U$ (Thm. A.1 in \cite{prozen}). 

An example of a symmetry is invariance of the zero-temperature cavity from Eq. (\ref{sp}) under bosonic rotations $V=e^{i\phi \ph}$ (with $\ph=\aa^\dg\aa$). This is an example of a continuous symmetry which does not stem from a conserved quantity in $\mreg$. Instead, this symmetry stems from the generator $\mathcal{N}$ of the corresponding $\mathcal{V} = e^{i\phi\mathcal{N}}$, which commutes with $\LL $. The generator acts as $\mathcal{N}(\rin) = \ph\rin-\rin\ph$ and its commutation with $\LL $ can be checked by writing both in the notation from Appendix \ref{ap1}. The block-diagonalization of $\LL $ stemming from this symmetry corresponds to equations of motion for matrix elements $\bra n|\rin|m\ket$ with $m-n=r$ being decoupled from those with $m-n\neq r$ (Eq. (6.1.6) in \cite{zoller_book}). This is also true for Examples~\ref{sec34} and \ref{sec35} and will be used to calculate conserved quantities. In this way, symmetries can help compartmentalize evolution of both states and operators.

Whenever a steady state is unique, any symmetry will also be a symmetry of the steady state. {Symmetries can thus be used to classify \cite{bardyn} or determine properties \cite{popkov2013} of classes of unique steady states.} For the example of the previous paragraph, the vacuum $|0\ket\bra0|$ is rotationally invariant under $V$. When $\mout$ is not one dimensional, symmetries will rotate $\mout$ into itself \cite{baum2}. Symmetries can thus be used to perform unitary rotations on the steady-state subspace. Finding global symmetries, i.e., all operators commuting with $\LL $, is often intuitive in physical systems and some conditions are given in Appendix \ref{ap1}. One can also use the brute-force approach of finding the null space of the commutator-operator $[\LL ,\cdot]$ described in the appendix of \cite{schirmer_symmetries}.

We briefly mention the existence of anti-commuting symmetries such as chiral \cite{bardyn} or parity-time \cite{prosen2012,prosen2012a} for dissipative dynamics. These can reveal symmetries in the spectrum of $\LL $ and $\LL ^\dg$ \cite{prosen2012}, similar to the spectrum of a chirally-symmetric Hamiltonian being symmetric around zero.

\subsection{Subspace symmetries}\label{sec23}

In case a global symmetry necessary to perform a desired operation on $\rout$ either does not exist or is not easily obtainable, one can perform that operation directly using a \textit{steady-state subspace symmetry} (also symmetry for stationarity \cite{baum2}). Subspace symmetries are all $\uout$ such that
\be\label{cond}
\uout^\dg \rout\uout\in\mout.
\ee 
In other words, after enough time has passed and a given \(\rin\) has asymptotically approached $\rout$, a subspace symmetry will commute with $\LL $ restricted to $\mout$:
\begin{equation}\label{subspace}
e^{\LL t}(\uout^{\dg}\rout \uout)=\uout^{\dg}e^{\LL t}(\rout)\uout.
\end{equation}
Global symmetries are also subspace symmetries. While global symmetries are usually closely related to $H$ and $\hi$, non-global subspace symmetries relate directly to the dimension and structure of $\mout$. In the general block-diagonal structure of $\rout$ [e.g. fifth structure from Eq. (\ref{exam})], blocks can be rotated within themselves via unitary operations and blocks of the same size can be exchanged with each other via discrete operations. Since conserved quantities determine the size of the blocks of $\rout$ and since the Lie algebra of subspace symmetries also depends on respective block size, {\textit{the number of conserved quantities is equal to the dimension of the Lie algebra of subspace symmetries.} Illustrating this for the fifth structure from Eq. (\ref{exam}), there are 5 conserved quantities and the Lie algebra of subspace symmetries is $u(2)\oplus u(1)$ (whose dimension is $2^2+1=5$).}

Subspace symmetries that lie exclusively in $\mout$ can be extended to $\mreg$ as long as they maintain the condition from Eq. (\ref{cond}). These operations are precisely what allows one to control decoherence-free subspaces (DFS) and noiseless subsystems (NS) in quantum computation. {Subspace symmetry generators can also be approximated with physically realizable operations \cite{cats}.} By duality, it is possible to have subspace symmetries of $\LL ^\dg$. While all global symmetries are subspace symmetries of both $\LL$ and $\LL ^\dg$, there is no guarantee that subspace symmetries of either $\LL $ or $\LL ^\dg$ are  global.

\subsection{Parity \& discrete rotations}\label{sec24}

Having omitted discrete symmetries ($U$ where $\phi$ takes discrete values), we expound on parity since it is the simplest discrete symmetry and it is a good starting point for the further examples in the paper. Eq. (\ref{dissym}) shows that if one can find a non-trivial operator that commutes with everything in $\LL $, then one is lucky to have found both a symmetry and a conserved quantity of a system. It turns out that systems with parity conservation necessarily have such an operator and parity can be thought of as a symmetry almost in the unitary sense of Eq. (\ref{symreg}):
\be
\begin{array}{c}
~~~~~~[P,H]=[P,\hi]=0~~~\forall l\\
\begin{array}{rcl}
\Neswarrow ~~&  & ~~~\Searrow\\
\dot{P}=\LL ^\dagger(P)=0~~ & \Rightarrow & ~~~[\mathcal{P},\LL ]=0
\end{array}
\end{array}
\ee 
In the above, $\mathcal{P}(\hi)=P\hi P$. The proof is simple. Assuming $\LL ^\dg(P)=0$, it is possible to construct conserved positive- and negative-parity projections $\Pi_\pm = \half (I \pm P)$, which in turn must commute with all operators in $\LL $ (Lemma 7 of \cite{baum2}). Therefore, $P$ must commute with everything as well.\hfill $\blacksquare$ 

As shown above, systems with parity conservation will always have a global parity symmetry. The converse is not true, e.g. symmetry under photon number parity $P=e^{i\pi \ph}$ of single photon loss in Eq. (\ref{sp}) does not imply that photon number parity is a conserved quantity. In general, any set of $d$ conserved projection operators will partition $\mreg$ into $d^2$ subspaces which will evolve independently under $\LL $ (Thm. 3 in \cite{baum2}), with at least $d$ of the subspaces having their own steady state. One can extend the proof above for any idempotent linear superposition of conserved quantities. For example, the oscillator rotation $e^{i\frac{2\pi}{d}\ph}$ generates $\mathbb{Z}_d$ and can be used to write the $d$ projection operators in Eq. (\ref{oscosc}). These projectors, by the above proposition, will commute with all operators in $\LL$.

While parity conservation is sufficient for the existence of multiple invariant subspaces (a quantum memory), it is not sufficient (Example~\ref{sec31}) or even necessary for the existence of steady-state coherences. In other words, steady-state coherences $\r_{\m\n}$ [fourth structure from Eq. (\ref{exam})] can exist with or without a discrete symmetry (e.g. Sec.~5.3 in \cite{baum2}). Both of these cases are demonstrated pictorially via the two types of $\rout$ below:
\be \label{par}
\left(\begin{array}{cccc}
\vspace{0.025in}\r_{00} & \r_{01} & \leftarrow & ~~\\
\vspace{0.025in}\r_{10} & \r_{11} & \leftarrow\\
\vspace{0.025in}\uparrow & \uparrow & \nwarrow\\
\vspace{0.025in}
\end{array}\right)
,
\left(\begin{array}{cc|cc}\r_{00} & \leftarrow & \r_{01} & \leftarrow\\
\uparrow & \nwarrow & \uparrow & \nwarrow\\
\hline \r_{10} & \leftarrow & \r_{11} & \leftarrow\\
\uparrow & \nwarrow & \uparrow & \nwarrow
\end{array}\right)
\ee
In the above list, arrows represent parts of the space which converge to $\r_{\m\n}$. In the left example, the entire space converges to a two-by-two $\mout$, symbolizing a system with no parity symmetry. In the right example of a system with parity symmetry, the full space is ``cut-up'' into four independent subspaces, each of which converges to a steady state/coherence. In general, the existence of coherences does not depend on how $\rin$ converges to $\mout$. The Examples presented below will all converge to $\mout$ in a way similar to the right example from Eq. (\ref{par}).

\section{Examples}\label{sec3}

We present four Examples of physical systems which do not have a unique steady state. We direct the interested reader to further examples from fermionic systems \cite{prosen2010} and multi-level atoms \cite{baum2, jakob2004}. Example \ref{sec31} is that of a single-qubit dephasing model (specific case of Sec.~3.8.3 in \cite{wisemanmilburn} or Sec.~10.3.3 of \cite{gregoratti2009}) -- the simplest version of an information-preserving structure. The next two Examples deal with two-qubit systems. Example~\ref{sec32} is taken from recent experimental work that stabilizes Bell states using trapped ions \cite{zoller_stabilizers}. In Example~\ref{sec33}, a Hamiltonian is added to the previous case. 

Examples \ref{sec34}-5 deal with {single-mode} $2$-photon \cite{Simaan1975, *Gilles1993, Simaan1978} and $d$-photon \cite{Voigt1980, *Zubairy1980, *Klimov2003} absorption, respectively. A sample calculation of $\rout$ is provided in Example~\ref{sec35} for $\rin$ being a coherent state. These prominent quantum optical systems \cite{loudon1975, *Agarwal1987, *Hach1994,*Dodonov1997} have been gaining interest from the quantum information \cite{coolforcats,zaki} and optomechanics \cite{nunnenkamp2012, *borkje2013} communities. Generalized versions of $2$-photon absorption have recently been investigated in the context of nano-mechanical \cite{voje2013} and superconducting qubit \cite{cats} systems.

\subsection{Single-qubit dephasing channel}\label{sec31}

Consider one qubit undergoing dephasing on two of the three axes of the Bloch sphere. In this case, there is one jump operator $F=Z$ in Eq. (\ref{lind}) and no Hamiltonian (with $Z$ representing the corresponding qubit Pauli matrix). The master equation simplifies to
\be
\LL(\rin)=2\left(Z\rin Z-\rin\right).
\ee
Picking the eigenbasis of $Z$, $Z|\m\ket=(-)^\m|\m\ket$ with $\m=0,1$, one can see that the states $M_\m=|\m\ket\bra \m|$ will be steady but the coherence $|0\ket\bra1|$ will not survive. The steady-state density matrix is then
\be
\rout=\lim_{t\rightarrow\infty}e^{\LL t}(\rin)=\r_0|0\ket\bra0|+\r_1|1\ket\bra1|
.\ee
Naturally, one expects the system to remember the initial $Z$-component of $\rin$. One can see that $\LL^\dg=\LL$ since the jump operator is Hermitian, so the conserved quantities $J_\m=M_\m=|\m\ket\bra\m|$. Letting $v_Z=\text{Tr}\{Z\rin\}$ and using the correspondence from Eq.~(\ref{rhoeq}), one indeed determines that the $Z$-component is preserved and $\r_\m=\half[1+(-)^\m v_Z]$.

\subsection{Two-qubit dissipation}\label{sec32}

In Ref. \cite{zoller_stabilizers}, an $\LL $ comprising two jump operators (which are closely related to stabilizer generators of qubit codes \cite{sarma2013}) will have a unique Bell state as its steady state. We study a system with one of those jump operators whose $\rout$ will be of the form of the fourth structure from Eq. (\ref{exam}), i.e., a DFS.

Let $\mreg$ be the space of matrices acting on the Hilbert space of two qubits. Take an $\LL $ from Eq. (\ref{lind}) with sole jump operator ($c$ in Box 1 of \cite{zoller_stabilizers})
\be\label{deff}
F= \half\left(I-Z_{1}Z_{2}\right)X_{2}
\ee
where $X_i,Y_i$, and $Z_i$ are usual Pauli matrices for the $i$th qubit. The steady-state space $\mout$ is spanned by $M_{\mu\nu}$ with $\m,\n=0,1$ and
\be\begin{aligned}
M_{00}&=\frac{1}{4}\left(I+Z_{1}\right)\left(I-Z_{2}\right)\\
M_{11}&=\frac{1}{4}\left(I-Z_{1}\right)\left(I+Z_{2}\right)\\
M_{01}&=\frac{1}{4}\left(X_{1}+iY_{1}\right)\left(X_{2}-iY_{2}\right)
.\end{aligned}\ee
Intuitively, $\mout$ is equivalent to the space spanned by $|\Psi_{p}\ket\bra\Psi_{q}|$ where $p,q=\pm$ and the Bell states $|\Psi_{\pm}\ket = \frac{1}{\sqrt{2}}(|01\ket\pm|10\ket)$. Also, $M_{10}=M_{01}^\dg$ and $M_{\m\m}$ sum up to the identity on $\mout$. The $M_{\m\n}$ are orthonormal in the sense that $\text{Tr}\{M_{\m\n}^\dg M_{\a\b}\}=\delta_{\m\a}\delta_{\n\b}$. One can check that $\LL (M_{\m\n})=0$. By duality, there must exist quantities $J_{\m\n}$ such that $\LL ^{\dg}(J_{\m\n})=0$. In this case, one can easily determine the conserved quantities by seeing that
\be\begin{aligned}
J_{00}&=\frac{1}{2}\left(I+Z_{1}\right)\\
J_{01}&=\frac{1}{2}\left(X_{1}+iY_{1}\right)X_{2}
\end{aligned}\ee
commute with $F$. We also have $J_{11} = I-J_{00}$ (since identity is always conserved) and $J_{10} = J_{01}^\dg$ (since $\{J_{\m\n}^\dg\}=\{J_{\m\n}\}$). Notice that since $J_{\m\n}$ commute with everything in $\LL $, unitary operators built out of them will be (global) symmetries of the system.
Both $J_{\m\n}$ and $M_{\m\n}$ form the Lie algebra $u(2)$. Finally, using the correspondence from Eq. (\ref{rhoeq}), the steady state $\rout\in\mout$ for initial state $\rin\in\mreg$ can be expressed as
\be\begin{aligned}
\rout&=\sum_{\m,\n=0}^{1}\r_{\m\n}M_{\m\n}\\
\r_{\mu\nu}&=\text{Tr}\{J_{\m\n}^\dg\rin\}
.\end{aligned}\ee

Notice that $Z_1$ is a parity operator, meaning that the analysis from Sec.~\ref{sec24} holds. We depict the scenario in the following matrix, written in the $\{|00\ket,|01\ket,|10\ket,|11\ket\}$ basis:
\be\label{parityst}
\left(\begin{array}{cc|cc}
1 & 0 & 0 & 1\\
0 & \text{\dashbox{3}(10,12){\ensuremath{1}}} & \text{\dashbox{3}(10,12){\ensuremath{1}}} & 0\\
\hline \vspace{-.15in}\\
0 & \text{\dashbox{3}(10,12){\ensuremath{1}}} & \text{\dashbox{3}(10,12){\ensuremath{1}}} & 0\\
1 & 0 & 0 & 1
\end{array}\right)
\ee
In the above, solid lines divide the space into four subspaces which evolve independently under $\LL $. In each subspace we have written the corresponding conserved quantity $J_{\m\n}$. Dashed boxes are around those parts of the respective subspaces that belong to $\mout$. 

As a final note, if one adds another jump operator $F_2 = \half\left(I-Z_{1}Z_{2}\right)Y_{2}$, then the two-by-two structure of $\mout$ will remain, but with $J_{01}=M_{01}$ being the new off-diagonal conserved quantity. This quantity now has one less non-zero entry, signalling that an $\LL $ with both jump operators will no longer preserve all information about the coherence between the other Bell states $|\Phi_+\ket$ and $|\Phi_-\ket$ [with $|\Phi_\pm\ket=\frac{1}{\sqrt{2}}(|00\ket\pm|11\ket)$].
  
\subsection{Driven two-qubit dissipation}\label{sec33}

The assumption that $\LL $ does not have purely imaginary eigenvalues does not mean that $\LL $ cannot have a Hamiltonian. We add $H=\o X_2$ (with real parameter $\o$) to Example~\ref{sec32} in order to be able to pump some steady-state populations into the matrix subspace spanned by the other set of Bell states $|\Phi_{\pm}\ket$. Notice that $H$ can also be introduced by letting $F\rightarrow F+i\o I$. The full evolution is now
\be\label{qubit}
\LL (\rin)=-i\o\left[X_{2},\rin\right]+2F\rin F^{\dg}-F^{\dg}F\rin-\rin F^{\dg}F
\ee
with $F$ defined in Eq. (\ref{deff}). All $J_{\m\n}$ from the previous Example commute with $H$, so the parity structure of Eq. (\ref{parityst}) remains and there will be steady-state coherences. However,  $[H,F]\neq 0$ and there is competition between drive out of and dissipation into the $|\Psi_{p}\ket\bra\Psi_{q}|$ subspace (with $p,q=\pm$). The old $M_{\m\n}$ are now modified to include parts of the space $|\Phi_{p}\ket\bra\Phi_{q}|$ and the overlapping spaces $|\Psi_{p}\ket\bra\Phi_{q}|$ and $|\Phi_{p}\ket\bra\Psi_{q}|$. The steady states and steady-state coherences $\bar{M}_{\m\n}$ for this Example are
\be\begin{aligned}
\bar{M}_{00}&=\frac{1}{\zeta}[M_{00}+\frac{\o}{2}\left(I+Z_{1}\right)\left(\o I+Y_{2}\right)]\\
\bar{M}_{11}&=\frac{1}{\zeta}[M_{11}+\frac{\o}{2}\left(I-Z_{1}\right)\left(\o I-Y_{2}\right)]\\
\bar{M}_{01}&=\frac{1}{\zeta}[M_{01}+\frac{\o}{2}\left(X_{1}+iY_{1}\right)\left(\o X_{2}-iZ_{2}\right)]
,\end{aligned}\ee
where $\bar{M}_{10}=\bar{M}^\dg_{01}$ and  normalization $\zeta=\sqrt{2\o^{4}+4\o^{2}+1}$. It is clear that the $\bar{M}_{\m\n}\rightarrow M_{\m\n}$ as $\o\rightarrow 0$. In the $\o\rightarrow\infty$ limit, the drive balances the dissipation and all $\bar{M}_{\m\n}$ are equally distributed over both Bell-state subspaces. All that is left to do is to normalize $J_{\m\n}$ to ensure that $\text{Tr}\{\bar{J}_{\m\n}^{\dg} \bar{M}_{\a\b}\} = \delta_{\m\a}\delta_{\n\b}$:
\be
\bar{J}_{\m\n}=\frac{\zeta}{2\o^{2}+1}J_{\m\n}
.\ee 

{Finally, noticing that $F+i\o I$ is factorizable reveals that this Example is a noiseless subsystem (NS).} From Sec.~\ref{sec1}, we know that the $\bar{M}_{\m\n}$ will share a common matrix that can be factored out via a unitary transformation. Applying the transformation
$U=\half\left[I+Z_{1}\left(X_{2}-I\right)+X_{2}\right]$ to $\bar{M}_{\m\n}$, one obtains 
\be\begin{aligned}
U\bar{M}_{\m\n}U&=|\m\ket\bra\n|\otimes\frac{1}{\zeta}\begin{pmatrix}1+\omega^{2} & i\omega\\
-i\omega & \omega^{2}
\end{pmatrix}\\
U\bar{J}_{\m\n}U&=|\m\ket\bra\n|\otimes\frac{\zeta}{2\o^{2}+1}\begin{pmatrix}1 & 0\\
0 & 1
\end{pmatrix}
.\end{aligned}\ee 
One can see that subspace symmetries on this transformed $\mout$ will simply be unitary combinations of $|\m\ket\bra\n|\otimes I$. Additionally, it is clear that the transformed $\bar{J}_{\m\n}$ form the Lie algebra $u(2)$ (up to a constant).

\subsection{{Single-mode} two-photon absorption}\label{sec34}

Consider bosonic systems with jump operator $F = \aa^2$ with $[\aa,\aa^\dg]=I$. While this system is infinite, one can successfully analyze them for finite energy using a large finite Fock space spanned by $\{|n\ket\}_{n=0}^N$ (where $N\gg1$) \cite{schirmer}. This case is highlighted because it is an infinite counterpart to Example~\ref{sec32} and has the same four-dimensional structure of $\mout$, with basis $M_{\m\n}=|\m\ket\bra\n|$ in Fock space (with $\m,\n=0,1$). The diagonal conserved quantities $J_{\m\m}$ correspond to projectors on the even and odd subspaces respectively:
\be
J_{\m\m}=\Pi_\m\equiv\sum_{n} |2n+\m\ket\bra2n+\m|.
\ee
One can construct the photon number parity,
\be
P=\Pi_0 - \Pi_1=(-)^{\ph},
\ee
which commutes with $\aa^2$. Therefore, $\mreg$ is once again split into four independent subspaces. The conserved quantity for the  off-diagonal subspace,
\be\label{j01}
J_{01}=\frac{(\ph-1)!!}{\ph!!}\Pi_{0}\aa,
\ee
where $m!!=m(m-2)!!$ is the double factorial \cite{dfac}, was first discovered by Simaan and Loudon in the note \cite{Simaan1978} which motivated this work. It is an example of a conserved quantity that does not commute with operators in $\LL $. One can obtain such quantities by first isolating the space where they exist and then applying $\LL ^\dg$. Due to the parity structure, we know that $J_{01}$ is off-diagonal in the sense that $J_{01}=\Pi_{0}J_{01}\Pi_{1}$. Furthermore, $J_{01}$ has to overlap with its corresponding steady-state coherence $|0\ket\bra1|$. With those two constraints and symmetry of $\LL $ under $V=e^{i\phi\ph}$ (see Sec.~\ref{sec22}), $J_{01}$ must consist only of elements of the form $|2n\ket\bra2n+1|$ with $n=0,1,\ldots$. Assuming a solution of the form $J_{01}=j(\ph) \Pi_0 \aa$ and plugging into $\LL ^\dg(J_{01})=0$ yields a recursion relation for $j(\ph)$, whose solution is Eq. (\ref{j01}). 

Physically, $J_{01}$ represents how the environment distinguishes components of $\rin$. It will preserve information only from elements $|2n\ket\bra2n+1|$ since, in that case, the same number of photon pairs is lost in relaxing to $|0\ket\bra1|$. In all other even-odd basis cases, e.g. $|2n\ket\bra2n-1|$, different numbers of photon pairs are lost ($n$ vs. $n-1$ pairs for the example).

\subsection{Single-mode $d$-photon absorption}\label{sec35}

As a generalization of Example~\ref{sec34}, let $d>0$ and
\be\label{photon}
\LL (\rin)=2\aa^{d}\rin \aa^{\dg d}-\aa^{\dg d}\aa^{d}\rin-\rin \aa^{\dg d}\aa^{d}
.\ee
The dynamics of these systems have been analytically solved for all time \cite{Voigt1980, *Zubairy1980, *Klimov2003, Simaan1975, *Gilles1993, Simaan1978}. However, the advantage of this analysis allows one to bypass that tedious algebra and obtain the steady state directly using conserved quantities. In related work, a system is presented which has not been solved but for which all conserved quantities have been analytically determined \cite{cats}.

Note that the $d=1$ case is simply Eq. (\ref{sp}), which has a unique steady state. For the general case, let
\be\label{oscosc}
\Pi_\m=\sum_{n} |dn+\m\ket\bra dn+\m|=\frac{1}{d}\sum_{\n=0}^{d-1} e^{i\frac{2\pi}{d}(\ph-\m)\n}
\ee
be $d$ different projectors with $\m,\n=0,1,...,d-1$. Noting the cyclic relationship among projection operators,
\be
\Pi_\m \aa = \aa\Pi_{(\m+1)\text{mod}{d}} = \Pi_\m\aa\Pi_{(\m+1)\text{mod}{d}},
\ee
one can see that $[\Pi_\m,\aa^d]=0$. According to Sec.~\ref{sec24}, the Fock space is then partitioned into $d^2$ subspaces, each evolving independently. We can thus write
\be
\rout=\sum_{\m,\n=0}^{d-1}\r_{\m\n}|\m\ket\bra\n|
\ee
with $\r_{\m\n}=\text{Tr}\{J_{\m\n}^\dg\rin\}$. Extending the recipe of the previous Example, there are $d^2$ conserved quantities
\be
J_{\m\n}=\frac{j_{\m\n}\left(\ph\right)}{\sqrt{(\n)_{\n-\m}}}\Pi_{\m}\aa^{\n-\m}
\ee
where the square-root is from normalization, $J_{\n\m}=J^\dg_{\m\n}$,
\be
j_{\m\n}\left(\ph\right)=\prod_{p=0}^{\frac{1}{d}\left(\ph-\m\right)-1}\frac{2\left(dp+\nu\right)_{\n-\m}}{\left(dp+\nu\right)_{\n-\m}+\left(dp+\n+d\right)_{\n-\m}},
\ee
and the falling factorial $\left(x\right)_{n}=x\left(x-1\right)...\left(x-n+1\right)$. Since $(x)_0=1$, the diagonal conserved quantities simplify to $J_{\m\m}=\Pi_\m$. Since $\sum_\m J_{\m\m}=I$, only $d^2-1$ quantities are independent. The off-diagonal quantity simplifies to Eq. (\ref{j01}) for $d=2$ and only the identity remains for $d=1$. The $J_{\m\n}$ are reducible into a direct sum of $u(d)$ Lie algebras. There exists one algebra for each value of $\ph$. In other words, $\sum_{\m,\n}J_{\m\n}$ forms an infinite block-diagonal matrix with blocks of length $d$, diagonal entries of 1, and off-diagonal entries depending on $j_{\m\n}(\ph)$. Finally, note that only the piece of $\rin$ that initially lives in a given subspace $(\Pi_{\m}\rin\Pi_{\n})$ will contribute to the  corresponding $\r_{\m\n}$ in $\rout$. 

As an example calculation, we determine $\rout$ when $\rin=|\a\ket \bra \a|$, a coherent state $\aa|\a\ket=\a|\a\ket$ with $\a\in\mathbb{C}$. Since a coherent state fills the entire Fock space, all subspaces have non-trivial evolutions and equilibrate to
\be\label{sum}
\r_{\m\n}=\frac{\a^{\star\n-\m}e^{-\left|\a\right|^{2}}}{\sqrt{(\n)_{\n-\m}}}\sum_{n=0}^{\infty}\frac{j_{\m\n}(dn+\m)}{(dn+\m)!}(\left|\a\right|^{2})^{dn+\m}.
\ee
{Since $P_n=j_{\m\n}(dn+\m)/(dn+\m)!$ are polynomials in $n$, $\r_{\m\n}$ are generalized hypergeometric functions whose arguments will be roots of $P_{n+1}/P_n$ \cite{zeilberger}. }The diagonal elements simplify if instead we express $\Pi_\m$ using the right-hand side of Eq. (\ref{oscosc}),
\be\label{sum2}
\r_{\m\m}=\frac{1}{d}\sum_{\nu=0}^{d-1}e^{-i\frac{2\pi}{d}\mu\nu}\exp\left[\left|\alpha\right|^{2}\left(e^{i\frac{2\pi}{d}\nu}-1\right)\right]
.\ee
In the large $|\a|$ limit, $\r_{\m\m}\rightarrow1/d$, distributing populations equally among the diagonal steady states and retaining no information about $\a$. {For $\m\neq\n$ in this limit, $\r_{\m\n}$ converges to a constant times $e^{-i\theta(\m-\n)}$, thus storing the phase $\theta=\arg(\a)$ of the initial coherent state for any $d$.}

Taking a look at specific cases, for $d=1$, Eq. (\ref{sum}) is just $\r_{00}=1$. For $d=2$, expressing in the $|\m\ket\bra\n|$ basis,
\be\nonumber
\rout=\begin{pmatrix}\,\half(1+e^{-2\left|\a\right|^{2}})\,\,\,\,\,\,\, & \a^{\star}e^{-\left|\a\right|^{2}}I_{0}(\left|\a\right|^{2})\,\\
\text{c.c.} & \half(1-e^{-2\left|\a\right|^{2}})
\end{pmatrix}
,\ee
where $I_0$ is the modified Bessel function of the first kind \cite{dlmf}. In the large $|\a|$ limit, $\r_{01}\rightarrow e^{-i\theta}/\sqrt{2\pi}$. For $d=3$:
\be\begin{aligned}
\r_{01}&={\textstyle \a^{\star}e^{-\left|\a\right|^{2}}\,_{0}F_{2}\left[\frac{2}{3},\frac{5}{6};(\frac{|\a|^{2}}{3})^{3}\right]\overset{_{|\a|\rightarrow\infty}}{\longrightarrow}e^{-i\theta}\frac{\Gamma(\frac{2}{3})\Gamma(\frac{5}{6})}{2\pi}}\\
\r_{12}&={\textstyle \a^{\star}e^{-\left|\a\right|^{2}}\frac{\left|\a\right|^{2}}{\sqrt{2}}\,_{0}F_{2}\left[\frac{7}{6},\frac{4}{3};(\frac{|\a|^{2}}{3})^{3}\right]\longrightarrow\frac{3}{\sqrt{2}}e^{-i\theta}\frac{\Gamma(\frac{7}{6})\Gamma(\frac{4}{3})}{2\pi}}\\
\r_{02}&={\textstyle \a^{\star2}e^{-\left|\a\right|^{2}}\frac{1}{\sqrt{2}}\,_{0}F_{2}\left[1+\frac{\sqrt{2}}{3}i,1-\frac{\sqrt{2}}{3}i;(\frac{|\a|^{2}}{3})^{3}\right]}\\
&\,\,\,\,\,\,\,\,\,\,\,\,\,\,\,\,\,\,\,\,\,\,\,\,\,\,\,\,\,\,\,\,\,\,\,\,\,\,\,\,\,\,\,\,\,\,\,\,\,\,\,\,\,\,\,\,\longrightarrow e^{-2i\theta}\frac{\text{csch}(\frac{\sqrt{2}}{3}\pi)}{2\sqrt{3}}
\nonumber.\end{aligned}\ee
In the above, $\Gamma$ is the Gamma function and $_q F_p$ is the generalized hypergeometric function \cite{dlmf}. In summary, Eqs. (\ref{sum}) and (\ref{sum2}) match the steady-state result of \cite{Simaan1975,*Gilles1993, Simaan1978} for $d=2$ and generalize it to arbitrary $d$.

{
\section{Characterization of steady-state and infinite-time density matrices}\label{sec4}
}

We are now ready to combine previous developments regarding the structure of $\rout$ (Thm. 7 in \cite{baumr} and Thm. 5 in \cite{robin}) with the work here regarding the specific coefficients $\r_{\m\n}$. Remembering Sec.~\ref{sec1}, $\rout$ is unitarily equivalent to a block-diagonal form with blocks indexed, say by $\kappa$. Each block will be of dimension $n_{\kappa}$ and each basis element in each block, $M_{\m\n}^{(\kappa)}$, will share an $m_{\kappa}$-dimensional factor density matrix $T^{(\kappa)}$ (with $n_\kappa,m_\kappa\geq 1$). The matrix $T^{(\kappa)}$ is factored out via a unitary transformation, so all operators in this section are written in a frame unitarily equivalent to the one in Eq. (\ref{lind}). According to the correspondence from Eq. (\ref{basis}), each entry in each block will depend on the expectation value of $\rin$ with a corresponding  conserved quantity $J_{\m\n}^{(\kappa)}$. The characterization of $\rout$ with no oscillating coherences is then
\be\begin{aligned}\label{class1}
\rout&=\bigoplus_{\kappa}\left[\sum_{\m,\n=1}^{n_{\kappa}}\r_{\m\n}^{(\kappa)}|\m\ket_{\kappa}\bra\n|\otimes T^{(\kappa)}\right]\\
\r_{\m\n}^{(\kappa)}&=\text{Tr}\{J_{\m\n}^{(\kappa)\dg}\rin\}\\
\LL(|\m\ket_\kappa\bra\n|\otimes T^{(\kappa)})&=0\\
\LL ^{\dg}(J_{\m\n}^{(\kappa)})&=0
\end{aligned}\ee
The steady states $M_{\m\n}^{(\kappa)}=|\m\ket_{\kappa}\bra\n|\otimes T^{(\kappa)}$ are such that $\text{Tr}\{T^{(\kappa)}\}=1$ (so they are no longer a normalized basis as in previous sections; $ \text{Tr}\{M_{\m\n}^{(\kappa)\dg} M_{\m\n}^{(\kappa)\dg}\}\neq1$). Therefore, $\sum_{\m,\kappa}\r^{(\kappa)}_{\m\m}=1$, $\r^{(\kappa)}_{\n\m}=\r^{(\kappa)\star}_{\m\n}$, and $\r^{(\kappa)}_{\m\m}\r^{(\kappa)}_{\n\n}\geq|\r^{(\kappa)}_{\m\n}|^{2}$. The $|\mu\ket_{\kappa} \bra \nu|$ means that the $|\m\ket\bra\n|$ basis is different for each block $\kappa$. Conserved quantities $J_{\m\n}^{(\kappa)}$ are organized such that 
\be
\text{Tr}\{J_{\m\n}^{(\kappa)\dg}|\sigma\ket_{\lambda}\bra\tau|\otimes T^{(\lambda)}\}=\delta_{\kappa\lambda}\delta_{\m\s}\delta_{\n\tau}
.\ee
Notice that the \textit{shape} of the preserved space is determined solely by the set of dimensions $n_\kappa$. The \textit{total capacity} (number of independent variables) is $\sum_\kappa n_\kappa^2 -1$. If one further wants to characterize \textit{how} information is preserved, i.e. the detailed structure of $\rout$, then knowledge of the nullspaces of $\LL $ (steady states and steady-state coherences) and $\LL ^\dg$ (conserved quantities) is sufficient \cite{robin}. Finally, if one wants to know \textit{what} information is preserved upon initialization of the system with some $\rin$, then one needs to evaluate the expectation values of the initial density matrix with all conserved quantities. 

We now relax the assumption of no oscillating coherences. In this case, one needs to consider the \textit{infinite-time density matrix} $\rho_{\infty}$ consisting of all pure imaginary eigenvectors of $\LL$. From Appendix \ref{ap3}, we see that oscillating coherences are induced by rotations on steady-state coherences, meaning that inclusion of oscillating coherences will move some of the zero eigenvalues from Fig.~\ref{f1} onto the non-zero parts of the imaginary axis. Those rotations are caused by a Hamiltonian $H_{\infty}$ (contained in $H$ from Eq. (\ref{lind})) acting on each block $\kappa$ (but not acting on $T^{(\kappa)}$),
\be\begin{aligned}
\r_{\infty}(t)=&~e^{-iH_{\infty}t}\rout e^{iH_{\infty}t}\\
H_{\infty}=&\bigoplus_{\kappa}\left[\sum_{\m=1}^{n_{\kappa}}E_{\m}^{(\kappa)}|\m\ket_{\kappa}\bra\m|\otimes I^{(\kappa)}\right]
,\end{aligned}\ee
where $I^{(\kappa)}$ are identity matrices on the respective spaces of $T^{(\kappa)}$. For ease of presentation, it is assumed that $|\m\ket_\kappa$ is the eigenbasis of $H_\infty$. Thus, $\r_{\m\n}^{(\kappa)}$ with $\m\neq\n$ may begin to rotate at frequencies $\lambda_{\m\n}^{(\kappa)}$ that consist of energy differences of $H_\infty$. Conserved quantities may no longer be conserved due to the induced rotations, so they are renamed $J\rightarrow S$. The complete characterization is then
\be\begin{aligned}\label{class2}
\rho_{\infty}(t)&=\bigoplus_{\kappa}\left[\sum_{\m,\n=1}^{n_{\kappa}}\r_{\m\n}^{(\kappa)}e^{i\lambda_{\m\n}^{(\kappa)}t}|\m\ket_\kappa\bra\n|\otimes T^{(\kappa)}\right]\\
\r_{\m\n}^{(\kappa)}&=\text{Tr}\{S_{\m\n}^{(\kappa)\dg}\rin\}\\
\LL(|\m\ket_\kappa\bra\n|\otimes T^{(\kappa)})&=i\l_{\m\n}^{(\kappa)}|\m\ket_\kappa\bra\n|\otimes T^{(\kappa)}\\
\LL ^{\dg}(S_{\m\n}^{(\kappa)})&=-i\lambda_{\m\n}^{(\kappa)}S_{\m\n}^{(\kappa)}\\
\lambda_{\m\n}^{(\kappa)}&=E_{\n}^{(\kappa)}-E_{\m}^{(\kappa)}\end{aligned}\ee
With this convention, it is easy to see that oscillating coherences can be removed by going into the rotating frame of $H_\infty$. The form of $H_\infty$, up to an arbitrary energy shift, can be obtained by determining all eigenvalues of $\LL $ lying on the imaginary axis.

The analogy with unitary evolution is strikingly straightforward. When there is no dissipation, given a $\rin$ and the eigenvalues/vectors of the super-operator $-i[H,\cdot]$, one can determine dynamics of the system for all time. Analogously, when dissipation is present, $\rin$ and selected eigenvalues/vectors of $\LL $ will determine the complete dynamics of the system after the dissipative behavior has subsided. In both cases, the relevant eigenvalues are only those on the imaginary line (unfilled circles in Fig. \ref{f1}). The additional complication of dissipation is that there will be two sets of eigenvectors, left and right, due to the lack of a Hermiticity condition on $\LL $.

\section{Discussion}\label{sec5}

\begin{figure}
\centering
\includegraphics[width=1.0\linewidth]{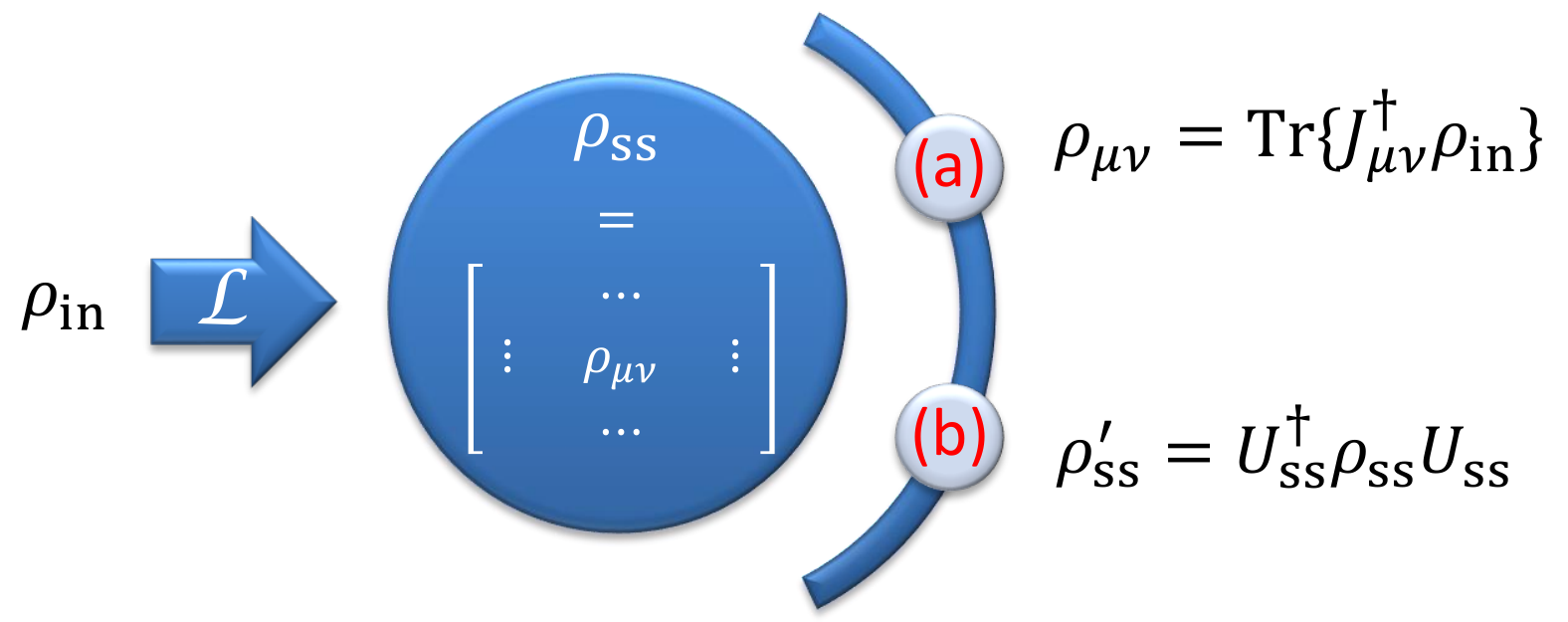}
\caption{(color online) Process of utilizing the steady-state subspace of a Lindblad system $\LL $ for quantum computation. Initial information is fed in via an initial density matrix $\rin$ which then equilibrates to a steady-state density matrix $\rout$ in the multi-dimensional steady-state space $\mout$. (a) The coefficients $\r_{\m\n}$ storing information about $\rin$ can be read out using conserved quantities $J_{\m\n}$. (b) Steady-state subspace symmetries $\uout$, conserved on $\mout$, can be to manipulate $\rout$. }
\label{f2}
\end{figure}

An often-discussed application of dissipative systems with many steady states is quantum information storage and computation. Unitary symmetries, both global and subspace, provide gates that can be performed on the steady state space without leaving the space. The capacity of $\mout$ as a computational space, i.e. noiseless code (and unitarily noiseless code in the case of $\r_\infty$ \cite{robin}), has been thoroughly studied \cite{tarasov, lidar2003, kempe2001}. Conserved quantities, on the other hand, can reveal how the information provided by an input state is stored in the output. This eliminates the need for any constraints on state initialization or the operators in $\LL $ \cite{shabani2005}, tracking dissipative evolution without error. Not all $\rin\notin\mout$ will completely lose all of their information -- an apt example of this is the encoding of the phase of a coherent state in $\rout$ using $d$-photon absorption from Example \ref{sec35}. Although difficult to physically interpret, Hermitian combinations of $J_{\m\n}$ are formal observables which can potentially be experimentally realizable. The application is illustrated in Fig. \ref{f2} (for one block), where the storage/readout of information is depicted in (a) and manipulation in (b). The only requirement for (a) is $\text{dim}\{\mout\}>1$, regardless of the dimension $n_\kappa$ of individual blocks, so $\mout$ does not have to be an NS/DFS for storage purposes. Note that this scheme is different from earlier work \cite{Verstraete2009} which utilized unique steady states to store information. In \cite{Verstraete2009}, the unique steady state is independent of $\rin$ and stores information about $\LL $. In this case, populations and coherences $\r_{\m\n}$ store information about $\rin$ and the information stored in the steady-state basis elements about $\LL $ is not used.

The above characterization also offers a slightly different framework for dealing with a subtle issue with DFS/NS. Strictly speaking, DFS's (Thm. 4 in \cite{lidar2003}) and NS's \cite{kempe2001} for Lindblad systems are defined in terms of \textit{only} the jump operators. While parts of $H$ can sometimes be absorbed into the jump operators (see Eq. (3.73) in \cite{breuer}), this may not always be the case. This ambiguity has spurred much work in determining different conditions on Hamiltonians $H$ in $\LL $ such that the respective DFS/NS is not broken \cite{kempe2001, shabani2005, brooke2008, Ticozzi2008}. In short, $H$ can have multiple functions, including determining $\mout$ (along with $\hi$) and inducing oscillating coherences on $\rout$. Since $H$ is included in the formulation presented here, one does not have to worry about putting any constraints on the Hamiltonian. If one wants to additionally manipulate $\rout$, one can generate subspace symmetries via control Hamiltonians $\hout(t)$ in the following way. First, let $\rin$ equilibrate to $\rout$. After that, we know that the two evolutions generated by $\LL $ and $\hout(t)$ commute [see Eq. (\ref{subspace})]. Thus, it is possible to combine the time-evolution generators into 
\be
\LL _\mathsf{ss}(\rout)=\LL (\rout)-i[\hout(t),\rout]
.\ee
Loosely speaking, if one can adiabaticaly turn-on $\hout(t)$ after $\rout$ has been formed, then one can use $\hout(t)$ to perform operations on $\rout$ with the only restriction being the condition on the unitary of $\hout$ from Eq. (\ref{cond}). {While exact gate Hamiltonians are guaranteed to exist due to subspace symmetries, they can also be approximated with realizable physical operations (e.g. in superconducting cavity systems \cite{cats}).}

Another area of investigation involving Lindblad generators is Markovian quantum feedback \cite{wiseman1993, wisemanmilburn, gregoratti2009} -- the study of using the output of an open quantum system to control its dynamics. Since experimental monitoring of a quantum system is often done almost continuously, much work has been done on simulating Lindblad master equations (or \textit{unraveling}) with \textit{stochastic master equations} (SME's). A typical stochastic master equation is a Lindblad equation with Gaussian noise terms for those $\hi$ which are deemed as the experiments' \textit{observed channels} (see Sec.~7.1.2 of \cite{gregoratti2009}). Therefore, the average evolution of all experimental runs (or numerical simulations of the SME) will follow the dynamics of the original Lindblad equation. However, individual simulations may not converge to exactly the Lindblad steady state. For example, when $\mout$ consists of two steady states and no steady-state coherences (e.g. the dephasing case of Ex.~\ref{sec31}), individual trajectories converge to one of the two possible states while the Lindblad steady state is a classical mixture of the two states and depends on initial conditions (see \cite{wangwiseman} for related simulations). Of course, conserved quantities can determine the averaged asymptotic state without having to run any simulations. Additionally, conserved quantities can help analyze the relative stability of various points in the steady-state space. While no state in a non-trivial $\mout$ is globally attractive (since one can end up in any steady state by simply starting in it; also see Cor.~1 in \cite{schirmer}), states in $\mout$ can have varying degrees of local attraction. Conserved quantities will help chart this attraction landscape. In other words, one can determine the values of parameters in $\rin$ that would make $\rout$ a specific pure state (\textit{the stable set} \cite{teschl2012}). Such analysis was done for a driven version of Example~\ref{sec34}, which exhibits two locally attractive pure steady states for some values of parameters in $\LL$ \cite{cats}.

It is worth discussing two physical requirements for the above applications to be realizable. First, one must wait for some time in order for information from $\rin$ to fully flow into $\mout$ (unless of course $\rin\in\mout$). Eigenvalues of $\LL $ make up the rates of exponential decay of states outside $\mout$ (which will be multiplied by powers of time $t$ in case of degeneracy; see Eq. (10.23) in \cite{puri}). The slowest rate of decay will be the eigenvalue of $\LL $ whose real part is closest to zero (the dissipation gap $\Delta$ from Fig. \ref{f1}). The dissipation gap will govern equilibration dynamics and should be \textit{large} compared to the timescale of the experiment. Second, the timescales from other sources of dissipation that destroy $\mout$ should be \textit{small} compared to the time one needs to store the information. For example, single photon loss in a cavity will always ruin any multi-dimensional $\mout$ (see Sec.~\ref{sec11}). The rates at which states approach $\mout$ and at which $\mout$ breaks down will depend on $\rin$, so not all $\rin$ may be practical even though the theory works for arbitrary initial state. One will have to determine how sensitive $\rin$ and states in $\mout$ are to all forms of malevolent decoherence present in the setup. Finally, note that even though $\mout$ is decoherence-free, it does not mean that $\rout$ will be a pure state since $|\r_{\m\n}|$ can vary between 0 and $\sqrt{\r_{\m\m}\r_{\n\n}}$, depending on $\rin$. However, whether $\rout$ is pure, slightly impure, or totally mixed, purity will be conserved under $\uout$.

\section{Outlook}\label{sec6}

{This work answers the question regarding what information is preserved when a density matrix evolves under a Lindblad generator. The paper provides a pedagogical explanation of the utility of symmetries and conserved quantities in dissipative Lindblad systems, showing that a symmetry-based analysis can be comparably as powerful and useful as it is with Hamiltonian systems despite the absence of a Noether-type theorem. We determine the utility of conserved quantities in obtaining the infinite-time density matrix. Analogous to unitary evolution, dissipative evolution in the infinite-time limit can be completely characterized by the initial state and purely imaginary eigenvalues of the Lindblad operator $\LL$. The major difference from unitary evolution is that there are two sets of eigenvectors -- those of $\LL$ and those of its adjoint -- due to non-Hermiticity of $\LL$. }

Practically speaking, conserved quantities and symmetries provide both an intuitive physical framework and a set of tools  to understand and manipulate steady-state density matrices carrying information. These tools should prove useful in theoretical formulations of decoherence-free subspaces/noiseless subsystems and further experimental developments of systems with multiple steady states.


\begin{acknowledgments}
Valuable discussions with Z. Leghtas, R. J. Schoelkopf, M. Pletyukhov, B. Bradlyn, Z. K. Minev, S. M. Girvin, S. Touzard, M. Mirrahimi, and M. H. Devoret are acknowledged. This work is supported by an NSF Graduate Research Fellowship,  the Alfred P. Sloan Foundation, DARPA Quiness program, and NBRPC (973 program) grant 2011CBA00300 (2011CBA00301).
\end{acknowledgments}

\appendix

\section{Notation and counter-examples}\label{ap1}

While in the text it is emphasized $M_\m$ are matrices, in the Appendices we switch to ``bra-ket'' notation for $\mreg$:
\be\begin{aligned}
\LL (\rin)&\longleftrightarrow \tl|\rin\kket \\
\text{Tr}\{J_{\m}^{\dg}M_{\n}\}&\longleftrightarrow \bbra J_{\m}|M_{\n}\kket\\
F \rin F^\dg&\longleftrightarrow(F\otimes F^\star)|\rin\kket.
\end{aligned}\ee
Note that the isomorphism induces an additional transposition on operators acting from the right on $\rin$ (see Sec.~2.1.4.5 of \cite{ernstnmr} for details). The ket $|\rin\kket$ is simply the $N$-by-$N$ $\rin$ written as an $N^2$-by-1 vector. In this equivalent form, $\tl$ is a matrix acting on $|\rin\kket$ from the left and is written as
\ba\label{matrix}
\tl&=&-i\left(H\otimes I-I\otimes H^{\star}\right)\nonumber\\
&&+\sum_{l=1}^{N^{2}-1}2\hi\otimes\hi^{\star}-\hi^{\dg}\hi\otimes I-I\otimes(\hi^{\dg}\hi)^\star
.\ea
The generator of a continuous symmetry $\mathcal{U}$ is written as
\be\label{jjj}
\mathcal{\hat{J}}=J\otimes I-I\otimes J^{\star}.
\ee
One can obtain conditions on $J$ for which $\mathcal{\hat{J}}$ is a symmetry generator from $[\mathcal{\hat{J}},\tl ]=0$:
\be\begin{aligned}
\sum_{l=1}^{N^{2}-1}[\hi,J]\otimes\hi^{\star}-\hi\otimes[\hi^{\star},J^{\star}]=0\\
\sum_{l=1}^{N^{2}-1}[\hi^{\dg}\hi,J]=[H,J]=0
.\end{aligned}\ee
If we assume one $\hi$, then the top condition becomes $\left[F,J\right]=\xi F$ with $\xi\in\mathbb{R}$. When $\xi\neq0$, the symmetry generates rotations on $F$. A simple example of a symmetry that neither commutes with $F$ nor is conserved is Eq. (\ref{sp}): $F=\aa$, $H=0$, and $J=\ph$. An example of a conserved quantity that neither commutes with $F$ nor is a symmetry is Example~\ref{sec34}: $F=\aa^2$, $H=0$, and $J=J_{01}+J_{01}^\dg$. Ref. \cite{baum2} provides other interesting examples.

\section{Proof of correspondence}\label{ap2}

Following Appendix \ref{ap1}, notation is continued.  The adjoint $\tl^{\dg}$ is defined for $|\r\kket,|\s\kket\in\mreg$ as 
\be
\bbra \s|\tl\r\kket=\bbra\tl^\dg \s|\r\kket
.\ee
Taking the adjoint of the Eq. (\ref{matrix}) obtains the bra-ket form of $\LL $ from Eq. (\ref{adj}). The adjoint has the same set of eigenvalues and eigenspaces of the same dimension as $\tl$. Therefore, there exist $D$ linearly independent conserved quantities $|J_{\mu}\kket$ such that $\bbra J_\m|\tl=0$. Since they are conserved, one can write
\be\label{jm}
\bbra J_{\m}|\rin\kket=\bbra J_{\m}|\rout\kket=\sum_{\n=1}^{D}\r_{\n}\bbra J_{\m}|M_{\n}\kket.
\ee
The matrix $\tl$ can be put into Jordan normal form via a non-unitary similarity transformation $S$,
\be\label{jordan}
\tl=S^{-1}\Lambda S.
\ee
Since the $|J_\m\kket$ and $|M_\m\kket$ are proper eigenvectors and there are no generalized eigenvectors (Lemma 17 of \cite{baum2}), the Jordan block with eigenvalue zero of $\Lambda$ will be simply a $D$-by-$D$ matrix of zeros. The respective transformed left and right eigenvectors, $|\tilde{M}_{\m}\kket = S|M_{\m}\kket$ and $\bbra\tilde{J}_{\m}| = \bbra J_{\m}|S^{-1}$, will be linearly independent and orthogonal to all other basis vectors of $\mreg$. Thus they are dual bases and can be made to be biorthogonal \cite{brink2001}, i.e., such that $\bbra\tilde{J}_{\m}|\tilde{M}_{\n}\kket=\delta_{\m\n}$. It is clear that once the transformed vectors are biorthogonal, the original ones are also: $\bbra J_{\m}|M_{\n}\kket=\delta_{\m\n}$. Plugging that into Eq. (\ref{jm}) obtains the desired result of Eq. (\ref{rhoeq}).\hfill $\blacksquare$

\section{Oscillating coherences}\label{ap3}

An oscillating coherence $|O\kket$ (also undamped oscillating phase relation \cite{baum2} or rotating point \cite{robin}) is an eigenvector of $\tl$ with a non-zero purely imaginary eigenvalue, i.e., 
\be
e^{\tl t}|O\kket=e^{i\l t}|O\kket
\ee
 for real $\l$. The presence of such an eigenvalue allows an initial density matrix to converge to a unitarily evolving state, i.e., a limit cycle \cite{schirmer, teschl2012} (also circular path \cite{baum1}). The source of oscillating coherences is an important non-trivial result from Thm. 18-3 of \cite{baum2}, where any pure imaginary non-zero eigenvalue is shown to stem only from the Hamiltonian part of $\tl$. In other words, if $H=0$, the limit set will consist entirely of steady states (with the converse being false; see Example~\ref{sec33}).

Another way to illustrate that oscillating coherences, steady states, and steady-state coherences form a complete basis for the limit set is by extending the proof from Appendix \ref{ap2} to eigenvalues on the imaginary axis. The fact that all such eigenvalues are proper (as opposed to generalized \cite{puri}) was proven in Lemma 2.3-ii of \cite{prosen2010} for quadratic fermionic $\tl $ and in Sec.~5 of \cite{schirmer} or Thm. 18 of \cite{baum2} for general Lindblad equations. The idea of the proofs is as follows. By contradiction, if one assumes that $\tl $ is not diagonalizable in the subspace of the Jordan normal form with diagonals of zero real part, then exponentiating the Jordan matrix of Eq. (\ref{jordan}), $e^{\tl t}=S^{-1}e^{\Lambda t}S$, will cause the dynamics to diverge as $t\rightarrow\infty$. For example:
\be\nonumber
\Lambda=\begin{pmatrix}i & 1\\
0 & i
\end{pmatrix}\longrightarrow e^{\Lambda t}=\begin{pmatrix}e^{it} & te^{it}\\
0 & e^{it}
\end{pmatrix}
.\ee

It is instrumental to think of oscillating coherences as rotating steady-state coherences. To illustrate this, let $\mout$ be $D$-dimensional with steady states $|M_{\m\m}\kket$ (with $\m=1,2,...,D$) and let $|O_{\m\n}\kket$ be oscillating coherences on all of the off-diagonals between the steady states, i.e., 
$e^{\tl t}|O_{\m\n}\kket=e^{i\l_{\m\n} t}|O_{\m\n}\kket$ for some real $\l_{\m\n}$. In the infinite time limit,
\be
|\r_\infty\kket=\sum_{\m=1}^D\r_{\m\m}|M_{\m\m}\kket+\sum_{\m\neq\n}\r_{\m\n}\left(t\right)|O_{\m\n}\kket.
\ee 
Unlike the unitary case, purity of $|\rin\kket$ may not be preserved: $|O_{\m\n}\kket$ collect information from $\rin$ just like $|M_{\m\m}\kket$, $\rho_{\m\n}(t)$ may depend on the structure of $\rin$, and $|\rho_{\m\n}(t)|\leq \sqrt{\r_{\m\m}\r_{\n\n}}$. The method to determine  $\rho_{\m\n}(t)$ is similar to the use of conserved quantities to determine $\r_{\m\n}$ from Eq. (\ref{rhoeq}), but this time the quantities $\bbra S_{\m\n}|$ corresponding to $|O_{\m\n}\kket$ are also rotating:
\be
e^{\tl^{\dg}t}|S_{\m\n}\kket=e^{-i\l_{\m\n}t}|S_{\m\n}\kket.
\ee 
These quantities are unique dual eigenvectors such that 
\be
\bbra S_{\m\n}|\r_\infty\kket=\rho_{\m\n}(t)
.\ee
However, it is also true that
\be
\bbra S_{\m\n}|\r_\infty\kket = \bbra S_{\m\n}|\lim_{t\rightarrow\infty}e^{\tl t}|\rin\kket=e^{i\l_{\m\n}t}\bbra S_{\m\n}|\rin\kket,
\ee
obtaining
\be
\rho_{\m\n}(t)=e^{i\l_{\m\n}t}\bbra S_{\m\n}|\rin\kket.
\ee
It is clear from the above that as $\l_{\m\n}\rightarrow 0$, $|O_{\m\n}\kket$ become steady-state coherences, $\bbra S_{\m\n}|$ become conserved quantities, $\rho_{\m\n}(t)$ stop rotating, and $|\r_\infty\kket\rightarrow|\rout\kket$. Since the $\l_{\m\n}$ stem from a part in the Hamiltonian of $\tl$, one way to eliminate oscillations is to go into the rotating frame of the generating Hamiltonian.

\linespread{1}    
\bibliographystyle{apsrev4-1}
\bibliography{D:/library} 

\end{document}